\documentstyle[prb,aps,array,multicol,epsfig]{revtex}
\tighten
\renewcommand{\theequation}{\arabic{section}.\arabic{equation}}
\newcommand{\ab}[1]{\setcounter{equation}{0}\section{#1}}
\newcommand{\gl}[1]{(\ref{#1})}
\newcommand{\ag}{\begin{equation}}
\newcommand{\eg}{\end{equation}}
\newcommand{\ega}{\end{eqnarray}}
\newcommand{\aga}{\begin{eqnarray}}

\begin{document}
\input epsf.tex
\draft 
\date{\today} 
\title
{Modified BCS mechanism of Cooper pair formation in narrow  energy bands of special symmetry\\
II. Matthias rule reconsidered}
\author{Ekkehard Kr\"uger}
\address{Max-Planck-Institut f\"ur Metallforschung, D-70506 Stuttgart, Germany}
\maketitle

\begin{abstract}
In part I of this paper a modified BCS mechanism of Cooper pair formation of electrons was proposed. This mechanism is connected with the existence of a narrow, roughly half-filled ``superconducting energy band" of given symmetry. The special symmetry of the superconducting band was interpreted within a nonadiabatic extension of the Heisenberg model of magnetism. Within this nonadiabatic Heisenberg model, the electrons of the superconducting band are {\em constrained} to form Cooper pairs at zero temperature because in any normal conducting state the conservation of crystal-spin angular momentum would be violated. Except for this participation of the angular momentum, the pair formation is mediated in the familiar way by phonons. Superconducting bands can be identified even within a free-electron band structure. Therefore, in this paper the band structures of the bcc and hcp solid solution alloys composed of transition elements are approximated by appropriate free-electron band structures with $s$-$d$ symmetry. From the free-electron bands, the number $n$ of valence electrons per atom related to the maxima of the superconducting transition temperature $T_{c}$ in these solid solutions is calculated within the nonadiabatic Heisenberg model. The observed two maxima of $T_{c}$ are reproduced without any adjustable parameter at valence numbers $n$ approximately equal to 4.9 and 6.5 in bcc and 4.7 and 6.7 in hcp solid solutions. This result is in good aggreement with the measured values of 4.7 and 6.4 of Hulm and Blaugher. The upper maximum is predicted not to exist in bcc transition elements but to occur in several ordered structures of bcc solid solution alloys.
\end{abstract}


\begin{multicols}{2}   
\narrowtext

\ab{Introduction}
\label{introduction}
In an early paper, Matthias\cite{ma} pointed out that the superconducting transition temperature $T_{c}$ depends crucially on the number $n$ of valence electrons per atom. Matthias proposed that $n = 5$ and $n = 7$ valence electrons per atom are particularly favorable for superconductivity. By means of this empirical rule, which is often referred to as Matthias rule, a great number of superconducting alloys composed of transition as well as nontransition elements have been discovered.

\begin{figure}[F1]  
\epsfxsize= 1 \hsize 
\centerline{ \epsffile{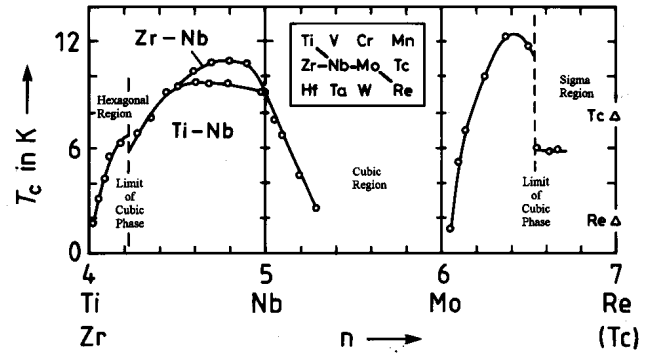}}
\vskip .5\baselineskip 
\caption{Transition temperature $T_{c}$ versus number $n$ of valence electrons per atom in solid solution alloys of the transition elements, as measured by Hulm and Blaugher (Ref.~\protect\onlinecite{hb}). The positions of the maxima are $n \approx 4.7$ and $n \approx 6.4$.}
\label{figure1}
\end{figure}

Later, Hulm and Blaugher\cite{hb} presented experimental data on the variation of $T_{c}$ with composition over the entire range of binary body-centered cubic solid solutions in the transition metal alloy system. They found two distinct maxima of $T_{c}$, one located between group 4 and group 5 close to the valence number $n \approx 4.7$, and the second located between group 6 and group 7 close to the valence number $n \approx 6.4$; see Fig.~\ref{figure1}. Though the positions of the observed maxima differ from the values of $n = 5$ and $n = 7$ originally proposed by Matthias, the results of Hulm and Blaugher confirm the essential physical substance of the Matthias rule, namely that $T_{c}$ strongly depends on the number $n$ of valence electrons per atom and that this $n$ dependence is similar in every alloy system.

As is well-known, there is overwhelming theoretical and experimental evidence that an electron-phonon mechanism is responsible for superconductivity in the transition elements.\cite{vo} In particular, the variation of $T_{c}$ with the number $n$ of valence electrons per atom is understood within the McMillan theory\cite{mc} of superconductors with strong electron-phonon coupling.\cite{go,va} Nevertheless, it cannot be excluded that this variation of $T_{c}$ can also be understood in a new way provided that this new way does not contradict the standard theory of superconductivity.

In this paper, the Matthias rule is reconsidered within the  nonadiabatic Heisenberg model\cite{es,ea,ef} which is defined by three new postulates given, e.g., in Ref.~\onlinecite{ef}. 

Niobium possesses a narrow, roughly half-filled energy band of special symmetry which in part I of this paper\cite{en} was called ``superconducting band" ($\sigma$ band). It was shown that {\em if} the postulates of the nonadiabatic Heisenberg model are satisfied within this $\sigma$ band, {\em then} the electrons of this band form Cooper pairs at zero temperature. Just as in the Bardeen-Cooper-Schrieffer (BCS) theory\cite{bcs}, the coupling of the electrons is mediated by phonons. However, the formation of Cooper pairs is constrained in an absolutely new way by the conservation law of spin angular momentum: within a narrow, partly filled $\sigma$ band the crystal-spin angular momentum is conserved at zero temperature if {\em and only if} the electrons form Cooper pairs.

Within the nonadiabatic Heisenberg model, the superconducting transition temperature $T_c^j$ of the $j$th $\sigma$ band of a metal may be calculated in the BCS limit from the modified BCS equation
\ag
T_c^j = 1.14\cdot\theta\cdot e^{-1/N^{\sigma}_j(E_F)V};
\label{bcssigma}
\eg 
see Sec.~IIIC of part I. $V$ and $\theta$ denote the effective electron-phonon interaction and the Debye temperature, respectively. The nonadiabatic Heisenberg model modifies the interpretation of the BCS formula in that the parameter $N^{\sigma}_j(E_{F})$ now is only a {\em partial} density of states of the electrons at the Fermi level, namely the density of states of the electrons of the $j$th $\sigma$ band.

In part I of this paper, only bcc metals with one atom per unit cell have been considered. The present paper, however, deals with materials with more than one atom in the unit cell. Therefore, single $\sigma$ bands must be replaced by ``$\sigma$-band complexes" containing just as many bands as there are atoms in the unit cell; see appendix~\ref{defsigmaband}.  
In equation~\gl{bcssigma} the index $j$ now labels the $j$th $\sigma$-band complex. In this paper, $j = L$ and $j = U$ denote the $\sigma$-band complex belonging to the lower and upper maximum of $T_{c}$, respectively. 

In Sec.~\ref{freeelectronbands} free-electron energy bands will be constructed which in this paper shall approximate the $s$-$d$ energy bands of the binary bcc and hcp solid solutions in the transition metal alloy system. That means, a free-electron band structure is constructed consisting only of linear combinations of plain waves with the symmetry of a $s$-$d$ band group. Because $\sigma$-band complexes can already be identified within these free-electron bands, in Sec.~\ref{calculation} the parameters $N^{\sigma}_{L}(E_{F})$ and $N^{\sigma}_{U}(E_{F})$ will be calculated within this free-electron model. With these $N^{\sigma}_{L,U}(E_{F})$, the {\em positions} of the maxima of $T_{c}^{L}$ and $T_{c}^{U}$ will be calculated by means of Eq.~\gl{bcssigma} without any fitting parameter, whereas the constant $V$ will be adjusted such that the maximum $T_{c}$ amounts to about $10K$. This adjustment does not influence the {\em positions} of the maxima noticeably.

In appendix~\ref{defsigmaband} the definition of a $\sigma$-band complex will be given in a metal with any given space group and any given number of atoms per unit cell, and in appendix~\ref{spindependentwfs} the related spin dependent Wannier functions (spin dependent Wfs) will be defined. 

\ab{Free-electron energy bands with \lowercase{s-d} symmetry}
\label{freeelectronbands}

The energy bands of the valence electrons of the transition elements are often referred to as $s$-$d$ bands since the valence electrons in isolated atoms of these metals have $s$ or $d$ symmetry. Consider a metal with the space group $G$ and $\mu$ atoms in the unit cell. In this paper, $6\mu$ energy bands of this metal are called ``$s$-$d$ bands" if, first, the Bloch functions of these bands can be unitarily transformed into Wfs centered at the atoms and if, secondly, at each atom there are six Wfs with one of them having $s$-like and the other five having $d$-like symmetry. $s$- and $d$-like means that in Eq. (1.8) of Ref.~\onlinecite{ew1} the Wfs belong to a representation of $G$ which is compatible with the representation $D_{0} + D_{2}$ of the three-dimensional rotation group $O(3)$. 

The band structure of the transition metals shall be approximated by a set of free-electron bands which satisfy the following three conditions.\\
(1) The set shall consist of $5\mu$ bands with the remaining $\mu$ bands connecting it with higher lying bands.\\
(2) Within the representation domain\cite{bc} of the Brillouin zone the energy $E_{i}(\vec k)$ of the $i$th band shall be calculated by
\ag
E_{i}(\vec k) = \displaystyle\frac{\hbar ^{2}}{2m}(\vec k - \vec K_{M})^{2},
\label{2.1}
\eg  
where $\vec K_{M}$ denotes the wave vector of the reciprocal lattice point $M$ related according to Table~\ref{table2} to the $i$th band.\\
(3) The bands shall be labeled by the representations given in Table~\ref{table1} and~\ref{table3}, respectively. Since, however, the set consists of $5\mu$ bands only, some representations may be absent.

\subsection{Body-centered cubic structure}

In the bcc structure we have $\mu = 1$. Table~\ref{table1} gives all the representations of a set of six $s$-$d$ bands in the Brillouin zone of the bcc lattice as derived from Table 2.7 of Ref.~\onlinecite{bc}.

In the band structure of the typical transition element Nb [the calculation of Mattheis\cite{mattheis} is depicted, e.g., in Fig.~1 of part I of this paper], the majority of the Bloch functions has the symmetry given in Table~\ref{table1}. However, Nb and the other transition elements do not possess an {\em ideal} set of $s$-$d$ bands because one level near the Fermi energy at point $N$ is labeled by the $p$-like representation $N_{1}'$.

The symmetry of the free-electron wave functions (which are symmetrized plane waves) can be determined by standard group theoretical methods.\cite{sl}

\begin{figure}[F2]  
\epsfxsize= 1.05 \hsize 
\centerline{ \epsffile{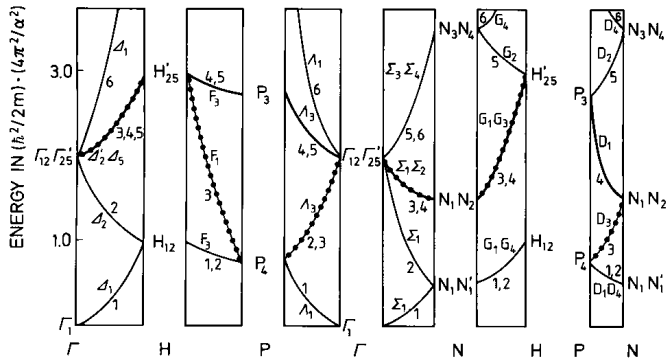}}
\caption{Free-electron $s$-$d$ energy bands which serve as approximation of the energy bands of the bcc solid solution alloys of the transition elements. The bands are numbered arbitrarily from 1 to 6 with the related points $M$ of the reciprocal lattice being given in Table~\ref{table2}a. Band 3 (dotted line) is a superconducting band. Band 4 (bold line) does not form a superconducting band, but it becomes a superconducting band complex when it is folded into the Brillouin zone of the MoSi$_{2}$ structure (see Fig~\ref{figure9}) or into the Brillouin zone of several atomic arrangements with the hexagonal space group $D^{3}_{3h}$.}
\label{figure3}
\end{figure}

It turns out that no set of free-electron bands exists which complies with all the three conditions and which satisfies the compatibility relations in the Brillouin zone. Therefore, as suggested by the band structures of the bcc transition elements, I replace one of the two $N_{1}$ levels by the $p$-like $N_{1}'$ level. With this modification of the third condition we get the one free-electron band structure given in Fig.~\ref{figure3} which is used in this paper as approximation for the energy bands of the bcc solid solution alloys of the transition elements. 

The symmetry structure of the free-electron bands has a close resemblance to what is found in the band structures of the bcc transition elements, whereas, of course, the free-electron energy $E_{i}(\vec k)$ deviates considerably from the actual energy. By ``symmetry structure" I mean the occurence and the connections of the representations between each other. For instance, in the free-electron case as well as in the actual bands, the $H_{12}$ state occures and is connected with the $P_{4}$ and the $N_{1}$ and $N_{1}'$ states.

\subsection{Hexagonal close-packed structure}
\label{sechcpstructure}
Table~\ref{table3} gives the representations of a set of $s$-$d$ bands in the hcp structure as determined by the methods derived in Refs.~\onlinecite{ew1} and~\onlinecite{ew2}. Now the set consists of twelve bands since there are $\mu = 2$ atoms in the unit cell. 

[The representations $R_{\vec k}$ given in Table~\ref{table3} satisfy Eq.~\gl{B9} if we put $d(\alpha ) = \chi_{0}(\alpha ) + \chi_{2}(\alpha )$, where $\chi_{0}(\alpha ) = 1$ and $\chi_{2}(\alpha )$ denote the characters of $\alpha$ in the representations $D_{0}$ and $D_{2}$, respectively, of the three-dimensional rotation group $O(3)$.]

\begin{figure}[F3]  
\epsfxsize= 1 \hsize 
\centerline{ \epsffile{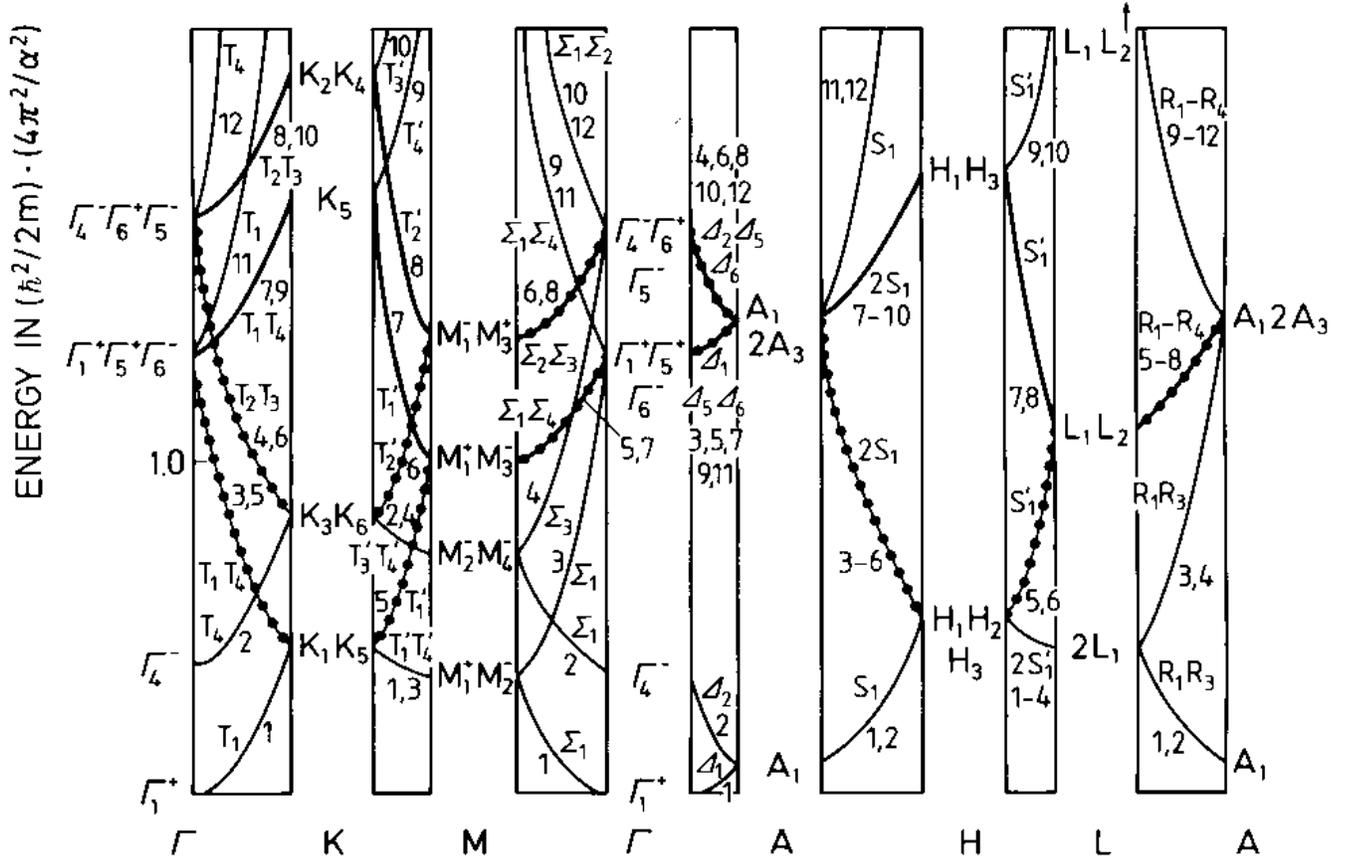}}
\caption{Free-electron $s$-$d$ energy bands which serve as approximation of the energy bands of the hcp solid solution alloys of the transition elements. The bands are numbered arbitrarily from 1 to 12 with the related points $M$ of the reciprocal lattice being given in Table~\ref{table2}b. Band 5 and 6 (dotted line) as well as band 7 and 8 (bold line) form superconducting band complexes of two bands each. Besides this, band 1 and 2 as well as band 3 and 4 form further superconducting band complexes not considered in this paper.}
\label{figure4}
\end{figure}

The free-electron band structure which can be constructed by starting from the three conditions at the top of this section, is depicted in Fig.~\ref{figure4}. Other sets of free-electron bands which also comply with these conditions, only differ unessentially from the band structure depicted in Fig.~\ref{figure4}. For instance, the upper $\Gamma^{+}_{1}$ and $\Gamma^{-}_{4}$ states may be interchanged. Hence, I use the band structure in Fig.~\ref{figure4} as approximation of the energy bands of the hcp solid solution alloys of the transition elements. 

In the case of the hcp structure, it is not necessary to use $p$-like states in the free-electron $s$-$d$ band structure in order for the compatibility relations to be fulfiled. In the real band structures of the hcp transition elements, however, $p$-like states are present and influence the symmetry-structure of the bands. In rhenium, e.g., a $p$-like state, namely $\Gamma^{+}_{3}$, is situated near the Fermi level; see Fig.~2 of Ref.~\onlinecite{mattheisr}. Therefore, an additional band exists in the band structure which causes a change of the symmetry structure. In this paper, this pertubation by the $p$-electrons will be ignored since it is difficult to include $p$-like states in an unequivocal way.

\ab{Calculation of the positions of the maxima of $T_c$.}
\label{calculation}

In this section I shall calculate the positions of the maxima of the superconducting transition temperature $T_{c}^{j}$ from equation~\gl{bcssigma}. The density of states $N^{\sigma}_{j}(E_{F})$ at the Fermi level is determined exactly for the free-electron bands given in Figs.~\ref{figure3} and~\ref{figure4}, the Debye temperature $\theta$ is arbitrarily put equal to $300K$ and the effective spin-phonon interaction $V$ is chosen in such a way that, in each case, the maximum transition temperature amounts to about $10K$. A variation of $V$ or $\theta$ does not modify the positions of the maxima of $T_{c}$ noticeably.

Changing the Fermi energy $E_{F}$ stepwise over the entire energy range of the considered set of $s$-$d$ bands, I calculate for each $E_{F}$ first the total number $n$ of electrons per atom by integration and then, for each $i$, the density of states $N_{i}(E_{F})$ of the $i$th band at the Fermi level. The calculated $n$ is referred to as the average number of valence electrons per atom. 

The density of states $N^{\sigma}_{j}(E_{F})$ is given by
\ag
N^{\sigma}_{j}(E_{F}) = \sum_{i} d_{ij}N_{i}(E_{F}),
\label{sumnsigma}
\eg
where 
\ag
d_{ij} = \left\{
\begin{array}{r l}
1 & \mbox{if the $i$th band belongs}\\
& \mbox{to the $j$th $\sigma$-band complex}\\
0 & \mbox{else.}
\end{array}\right.
\eg
In the following, we put $j = U$ and $j = L$ for the upper and lower maximum of $T_{c}$, respectively.

\end{multicols}
\widetext

\begin{multicols}{2}
\narrowtext

\subsection{Body-centered cubic structure}

For an integer number $n$ of valence electrons, the bcc solid solution alloys studied by Hulm and Blaugher\cite{hb} are elements with the space group $G = O_{h}^{9}$ and one atom per unit cell. $\sigma$-band complexes in this structure are simple $\sigma$ bands consisting of only one energy band.   

If $n$ is not integer, the space group of the alloy is a subgroup $G'$ of $O_{h}^{9}$. An energy band recognized as $\sigma$ band in $O_{h}^{9}$ stays a $\sigma$-band (complex) in the subgroup $G'$, i.e., it stays a $\sigma$-band (complex) when the symmetry is lowered on the fixed atomic bcc lattice. On the other hand, an energy band not forming a $\sigma$ band in $O_{h}^{9}$ may become a $\sigma$-band complex when it is folded into the Brillouin zone of the subgroup $G'$.

\subsubsection{Lower maximum of $T_{c}$}
 
Table~\ref{table4} lists the representations of all the $\sigma$ bands in the bcc structure. These representations are so-called extra or double-valued representations of the group $G_{\vec k}$ of the wave vector $\vec k$.

The only $\sigma$ band which can be found among the energy bands in Fig.~\ref{figure3} is band 3 denoted by the dotted line. This $\sigma$ band can be identified by means of a compatibility table for double-valued and single-valued representations. From the tables in Ref.~\onlinecite{elliott} we get
\aga
D_{1/2}\times\Gamma_{25}' &=& \Gamma^{+}_{7} + \Gamma^{+}_{8},\nonumber\\
D_{1/2}\times H_{25}' &=& H^{+}_{7} + H^{+}_{8},\nonumber\\
D_{1/2}\times N_{2} &=& N^{+}_{5},\nonumber\\
D_{1/2}\times P_{4} &=& P_{7} + P_{8}.
\label{3.2}
\ega
These equations show that the spin dependent Bloch functions $\phi_{\vec kqm}(\vec r,t)$ [given in Eq.~\gl{C6}] of a band with the single-valued representations 
\ag
\Gamma_{25}',\ H_{25}',\ P_{4},\mbox{ and }N_{2} 
\label{svrep}
\eg
can be transformed in such a way that at each symmetry point two functions form a basis of the double-valued representations
\ag
\Gamma^{+}_{7},\ H^{+}_{7},\ P_{7},\mbox{ and }N^{+}_{5}
\label{3.3}
\eg
given in the second line of Table~\ref{table4}. Band 3 in Fig.~\ref{figure3} is a $\sigma$ band because the representations~\gl{svrep} can be found among the representations related to band 3.

\begin{figure}[F4a]  
\epsfxsize= 1 \hsize 
\centerline{ \epsffile{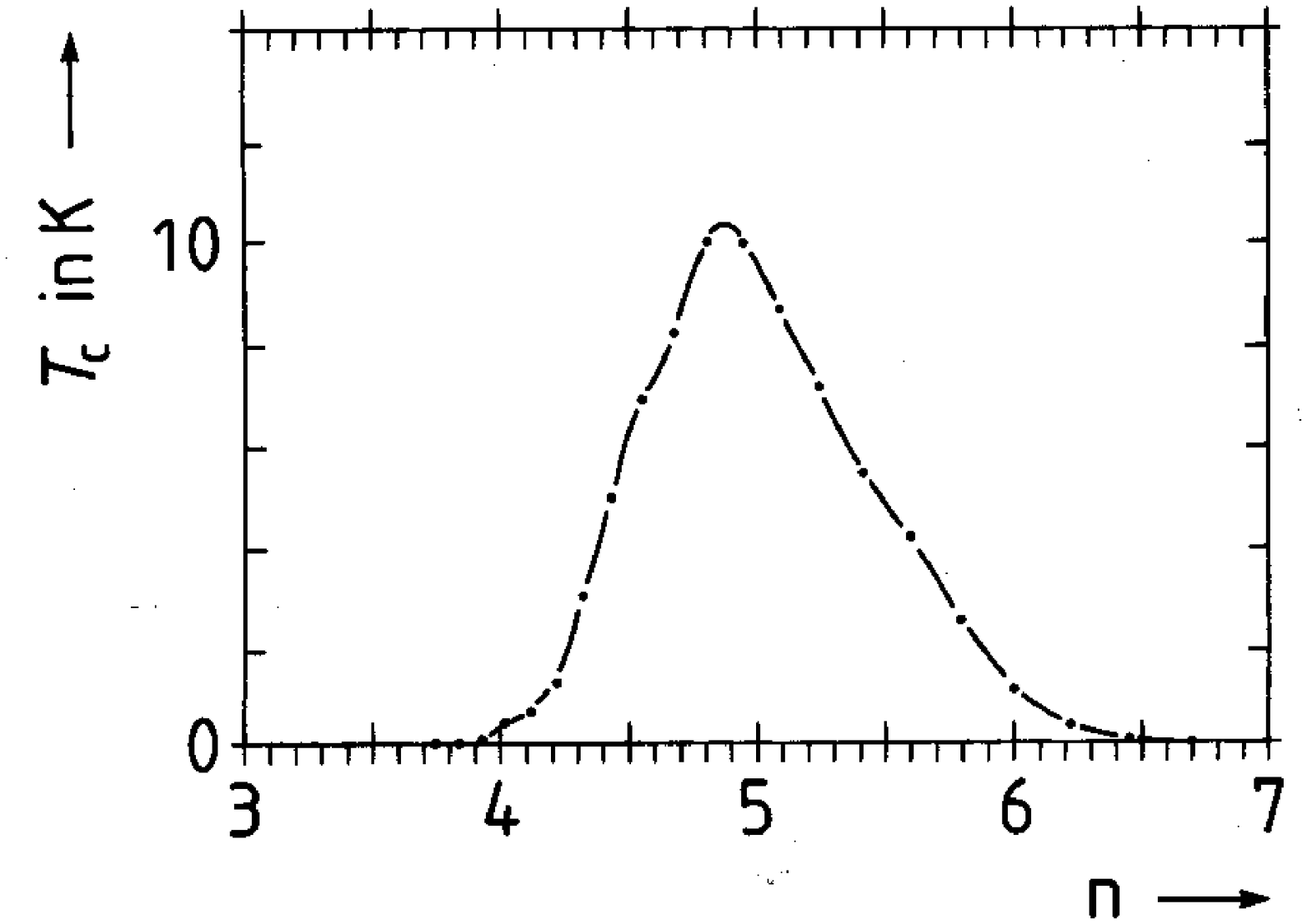}}
\vskip .5\baselineskip 
\caption{Transition temperature $T_{c}$ versus number $n$ of valence electrons per atom in bcc solid sulution alloys of the transition elements, as calculated from Eq.~\protect\gl{bcssigma} within the free-electron model by adjusting $V$ such that the maximum $T_{c}$ amounts to about $10K$. The positions of the maxima are independent of $V$.\\
Lower maximum at $n \approx 4.9$ related to the superconducting band 3 denoted by the dotted line in Fig.~1 of part I and Fig.~\ref{figure3}.}
\label{figure6a}
\end{figure}  

Fig.~\ref{figure6a} shows $T_{c}^{L}$, as calculated with
\ag
N^{\sigma}_{L}(E_{F}) = N_{3}(E_{F}),
\label{3.4}
\eg
as function of the number $n$ of valence electrons per atom. The maximum at $n \approx 4.9$ coincides with surprising accuracy with the experimental value $n \approx 4.7$ of Hulm and Blaugher.\cite{hb} Obviously, the $T_{c}(n)$ curve is determined to a large extend by the symmetry structure of the $s$-$d$ bands rather than by the real values of the energy $E(\vec k)$. This idea is corroborated by the calculated band structure of Nb [see Fig.~1 of part I], showing that a band with the symmetry~\gl{svrep} is roughly half-filled. This band has already been considered in Fig.~1 of Ref.~\onlinecite{es}.

\subsubsection{Upper maximum of $T_{c}$}

The upper maximum of $T_{c}$ is not present in the bcc structure. Band 4, as denoted in Fig.~\ref{figure3} by the bold line, is characterized by the (double-valued) representations
\ag
\Gamma^{+}_{7},\ H^{+}_{7},\ P_{8},\mbox{ and }N^{+}_{5},
\label{3.5}
\eg
indicating that at point $P$ this band is connected with other bands, since the representation $D_{1/2}\times P_{3} = P_{8}$ is four-dimensional. Therefore, $P_{8}$ cannot be found in Table~\ref{table4} and, hence, band 4 does not form a $\sigma$ band. This result, too, is confirmed by Hulm and Blaugher since the curve of the upper maximum does not go back to the integer value $n = 6$; see Fig.~\ref{figure1}. Later measurements have shown that both Mo and W (having six valence electrons per atom) become superconducting at the relatively low temperatures $0.92K$ and $0.0012K$, respectively. Their superconducting state is clearly related to band 3 because both metals have a band with the symmetry~\gl{svrep} in their band structure which is far from half, but not completely filled; see Fig.~1 in Ref.~\onlinecite{es2}.

\begin{figure}[F4b]  
\epsfxsize= 1 \hsize 
\centerline{ \epsffile{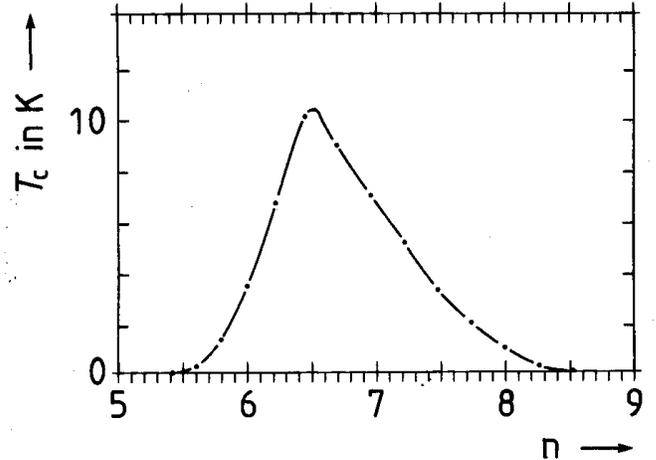}}
\vskip .5\baselineskip 
\caption{As Fig.~\ref{figure6a}, but the upper maximum at $n \approx 6.5$ related to band 4 denoted by the bold line in Fig.~\ref{figure3}. Because band 4 is not a superconducting band within the bcc structure, this maximum does not exist in the bcc transition elements, while it exists in some ordered structures in bcc solid solution alloys.}
\label{figure6b}
\end{figure}  

Fig.~\ref{figure6b} shows a plot of $T_{c}^{U}(n)$ as calculated with
\ag
N^{\sigma}_{U}(E_{F}) = N_{4}(E_{F}),
\label{nband6}
\eg
e.g., this curve is calculated by assuming that, nevertheless, band 4 is a $\sigma$ band. The calculated maximum of $T_{c}^{U}(n)$ at $n \approx 6.5$ coincides with the experimental value $n \approx 6.4$ with high accuracy. This fact together with the steep rise of the experimental $T_{c}(n)$ curve for $n > 6$ in Fig.~\ref{figure1} suggests that band 4 becomes a $\sigma$-band complex when the ideal bcc symmetry of the electron system is disturbed, i.e., when band 4 is folded into the Brillouin zone of the space group $G'$ of the Mo-Re or W-Re alloy.

Unfortunately, in the literature I have not found any details of the symmetry of this alloy in the bcc solid solution range. Possibly, the atoms are completely irregularly distributed over the bcc lattice sites. In this case, band 4 becomes clearly a $\sigma$ band since the problem with the $P_{3}$ state disappears. More likely, however, the atoms in these alloys are ordered or have at least a short-range order. Therefore, I have examined band 4 with the symmetry~\gl{3.5} in four ordered structures in a body-centered cubic solid solution.

\begin{figure}[F5]  
\epsfxsize= 1 \hsize 
\centerline{ \epsffile{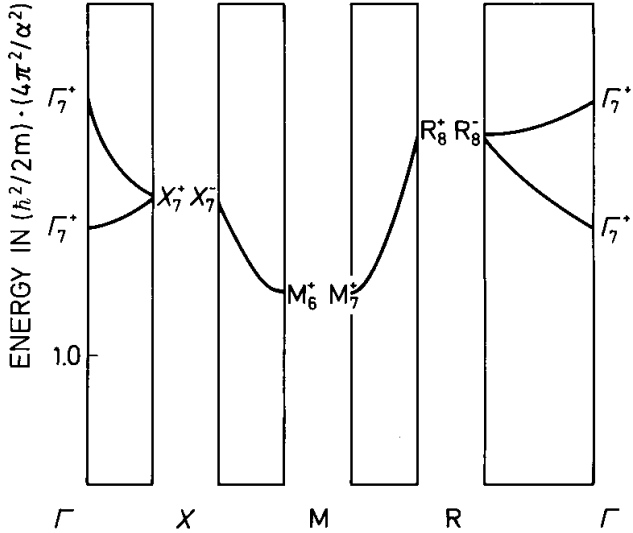}}
\vskip .5\baselineskip 
\caption{Band 4 in Fig.~\ref{figure3} folded into the Brillouin zone of the CsCl structure, with symmetry notations as used in Table~\ref{table5}. In this structure, band 4 does not form a superconducting band complex (cf. Table~\ref{table5}).}
\label{figure7}
\end{figure}  

First, consider the CsCl structure with the cubic space group $O_{h}^{1}$ and two atoms per unit cell. The double-valued representations of the four possible $\sigma$-band complexes (of two bands each) in the CsCl structure are listed in Table~\ref{table5}. Fig.~\ref{figure7} shows band 4 folded into the Brillouion zone of $O^{1}_{h}$. This set of two bands does not form a $\sigma$-band complex because the representations at point $R$, namely $R^{\pm}_{8}$, cannot be found in Table~\ref{table5}. The same result we get for the NaTl structure with the space group $O_{h}^{7}$. 
Consequently, we can exclude that the CsCl as well as the NaTl structure is realized in bcc solid solutions of Mo-Re or W-Re since in both structures the measured upper maximum of $T_{c}$ cannot exist.

\begin{figure}[F6]  
\epsfxsize= 1 \hsize 
\centerline{ \epsffile{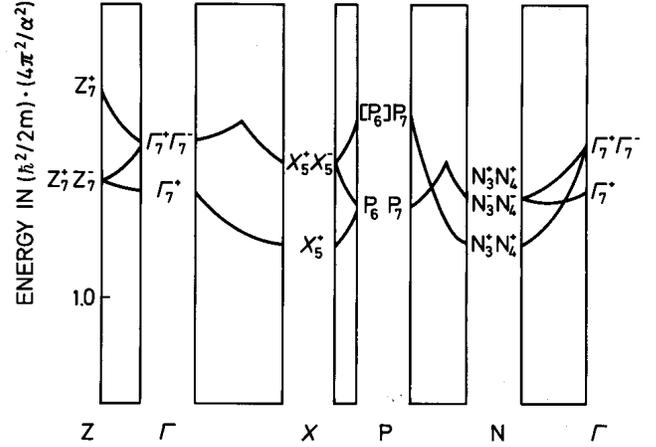}}
\vskip .5\baselineskip 
\caption{Band 4 in Fig.~\ref{figure3} folded into the Brillouin zone of the MoSi$_{2}$ structure, with symmetry notations as used in Table~\ref{table6}. According to Table~\ref{table6}, band 4 forms a superconducting band complex in this structure.}
\label{figure9}
\end{figure}

The MoSi$_{2}$ structure with the tetragonal space group $D^{17}_{4h}$ has been observed rather frequently, but often with distortions of the atomic bcc lattice.\cite{du} Table~\ref{table6} lists the (double-valued) representations of the four possible $\sigma$-band complexes in the MoSi$_{2}$ structure and Fig.~\ref{figure9} shows band 4 folded into the Brillouin zone of $D^{17}_{4h}$. The resulting set of three bands forms a $\sigma$-band complex as indicated by the second row of Table~\ref{table6}.

Other conceivable structures on an atomic bcc lattice have the hexagonal space group $D^{3}_{3d}$, as, e.g., the CdI$_{2}$ structure. [The International Tables give $\pm (\frac{1}{3},\frac{2}{3},z)$ for the positions of the I atoms. The atomic lattice has bcc structure for $z = \frac{2}{3}$.] A hexagonal space group is suggested by the neighbouring ``pseudo hexagonal"\cite{bergman} $\sigma$ structure (see Fig.~\ref{figure1}).  It turns out that, as in the MoSi$_{2}$ structure, band 4 becomes a $\sigma$-band complex in the CdI$_{2}$ structure when it is folded into the Brillouin zone of $D^{3}_{3d}$. 

To summarize, in all the considered structures on an atomic bcc lattice, except for the CsCl and NaTl structure, the upper maximum of the transition temperature as depicted in Fig.~\ref{figure6b} may exist.

\subsection{Hexagonal close-packed structure}

\begin{figure}[F7a]  
\epsfxsize= 1 \hsize 
\centerline{ \epsffile{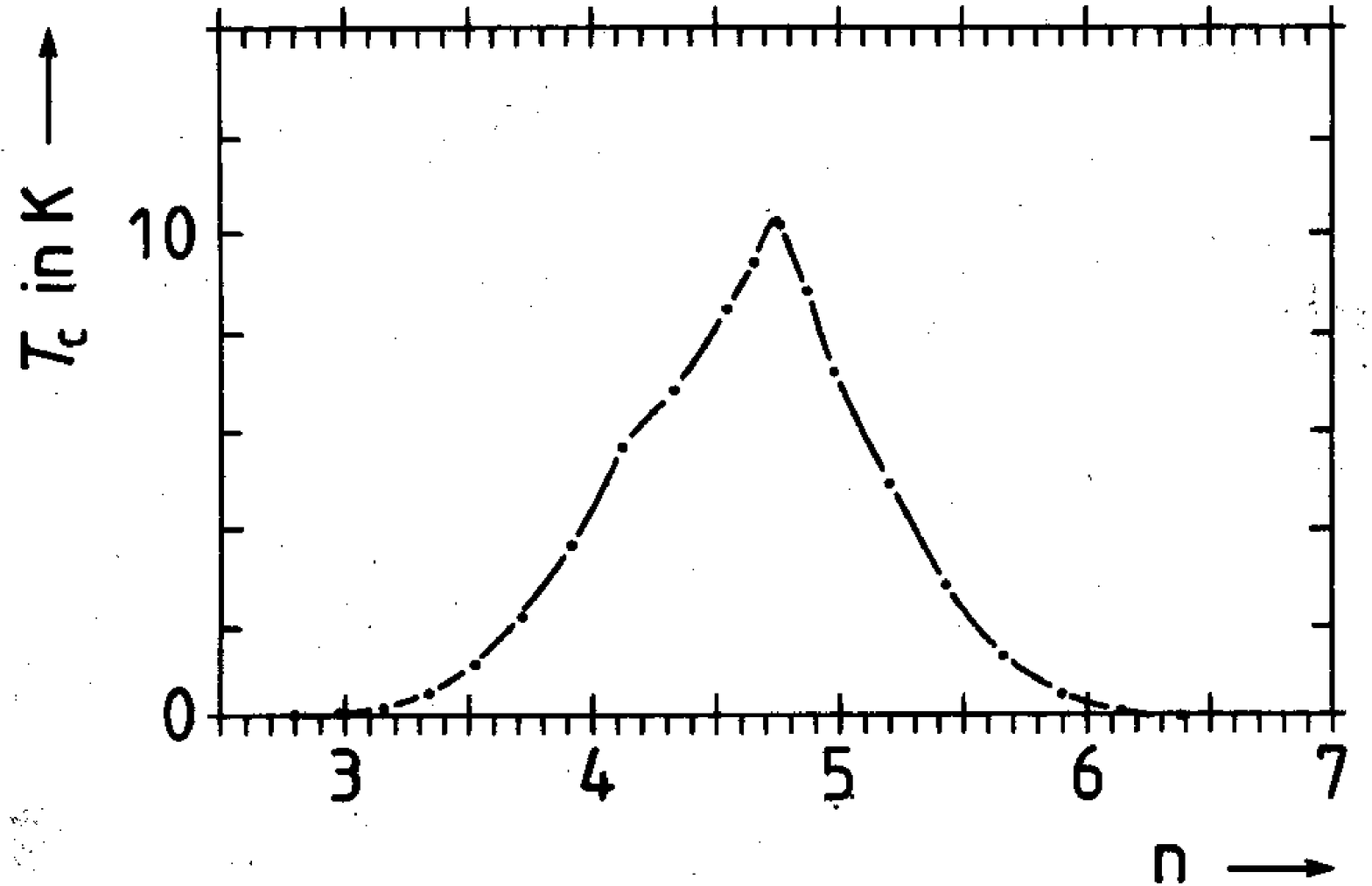}}
\vskip .5\baselineskip 
\caption{Transition temperature $T_{c}$ versus number $n$ of valence electrons per atom in hcp solid sulution alloys of the transition elements, as calculated from Eq.~\protect\gl{bcssigma} within the free-electron model by adjusting $V$ such that the maximum $T_{c}$ amounts about to $10K$. The positions of the maxima are independent of $V$.\\
Lower maximum at $n \approx 4.7$ related to the superconducting band complex consisting of band 5 and 6 which in Fig.~\ref{figure4} is denoted by the dotted line.}
\label{figure11a}
\end{figure}

\begin{figure}[F7b]  
\epsfxsize= 1 \hsize 
\centerline{ \epsffile{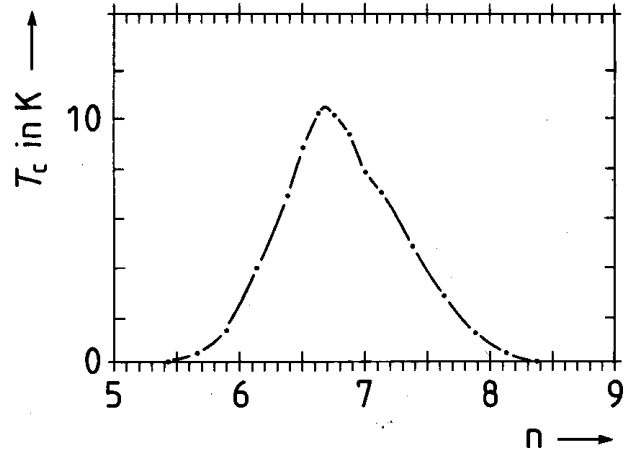}}
\vskip .5\baselineskip 
\caption{As Fig.~\ref{figure11a}, but the upper maximum at $n \approx 6.7$ related to the superconducting band complex consisting of band 7 and 8 which in Fig.~\ref{figure4} is denoted by the bold line.}
\label{figure11b}
\end{figure}

In the hcp structure we have two atoms in the unit cell and, hence, any $\sigma$-band complex consists of two energy bands. All the possible $\sigma$-band complexes in this structure are listed in Table~\ref{table7}. By means of the compatibility tables in Ref.~\onlinecite{elliott} it can be verified that the set of band 5 and 6 as well as the set of band 7 and 8 denoted in Fig.~\ref{figure4} by the dotted and bold lines, respectively, form $\sigma$-band complexes.\cite{misprint} Figs.~\ref{figure11a} and~\ref{figure11b} show $T^{L}_{c}$ and $T^{U}_{c}$, respectively, as function of $n$ calculated by Eq.~\gl{bcssigma} with
\ag
N^{\sigma}_{L}(E_{F}) = N_{5}(E_{F}) + N_{6}(E_{F}) 
\label{3.7}
\eg
and
\ag
N^{\sigma}_{U}(E_{F}) = N_{7}(E_{F}) + N_{8}(E_{F}).
\label{3.8}
\eg
Again the two maxima of $T_{c}$ at $n \approx 4.7$ and $n \approx 6.7$ coincide almost exactly with the measured values of $n \approx 4.7$ and $n \approx 6.4$, respectively. It should be noted that other possible definitions of the single free-electron bands yield slightly shifted positions of the maxima. This paper defines the bands as given in Eq.~\gl{2.1}.

\ab{Discussion}
Within the nonadiabatic Heisenberg model, the electrons of a narrow $\sigma$-band complex are constrained to form Copper pairs at zero temperature because in any normal conducting state the conservation of crystal-spin angular momentum would be violated; see part I of this paper. On the basis of this modified BCS mechanism of Cooper pair formation, a theoretical interpretation of the precise measurements of Hulm and Blaugher~\cite{hb}, see Fig.~\ref{figure1}, becomes possible within an appropriate free-electron model and within the BCS theory. 

Hence, there is great evidence that the BCS theory yields the {\em absolute} energy minimum in the Hilbert space only if the formation of Cooper pairs is additionally constrained by the constraining forces existing in a narrow, partly filled $\sigma$-band complex. Under this assumption, relevant results (but, of course, not all the results) of the strong coupling theory of superconductivity can already be obtained within the weak coupling (BCS) limit. Within the approximations of this paper, I was able to calculate the numbers $n$ of valence electrons per atom related to the maxima of the transition temperature $T_c$, but not the values of $T_c$.

The calculated positions of the maxima of $T_{c}$ at valence numbers approximately equal to 4.9 and 6.5 in bcc solid solution (see Figs.~\ref{figure6a} and~\ref{figure6b}) are in good agreement with the measured values of 4.7 and 6.4 by Hulm and Blaugher; see Fig.~\ref{figure1}. Obviously, the positions of the maxima of $T_{c}$ are determined to a large extend by the symmetry structure of the $s$-$d$ electrons rather than by the special form of the periodic potential of the ion cores. This theoretical result emphasizes the universal character of the Matthias rule.

In order to test the influence of the atomic structure on the transition temperature, the related maxima have been calculated in hcp solid solutions, too. Here we get nearly the same positions as in bcc solid solutions, namely $n \approx 4.7$ and $n \approx 6.7$; see Figs.~\ref{figure11a} and~\ref{figure11b}. This result suggests that the Matthias rule reflects a property of $s$-$d$ electrons which is not influenced much by the lattice symmetry. 

This statement, however, should be modified: A too high symmetry of the lattice, i.e., of the periodic potential, may suppress superconductivity. A remarkable finding is that the upper maximum of $T_{c}$ at $n \approx 6.5$ exists neither in the transition elements with bcc structure nor in alloys with CsCl or NaTl structure, while it may occur in several ordered structures in a bcc solid solution alloy of two components. These are, e.g., the tetragonal MoSi$_{2}$ structure and several atomic arrangements with the hexagonal space group $D^{3}_{3h}$. Obviously, this fact is responsible for the abrupt increase of the $T_{c}$ curve in Fig.~\ref{figure1} between $n = 6$ and $n = 6.3$.

The upper maximum in Fig.~\ref{figure1} is markedly narrower than the lower. This might indicate that a (short-range) order with the composition MoRe$_{2}$ with $n \approx 6.33$ is present in the solid solution. I believe that the ordered structure has the highest $T_{c}$, though, at this stage, I cannot give a theoretical foundation for this speculation.

The actual band structures of the transition metals contain also $p$-like Bloch states. In this paper I have taken into account $p$-like states in the bcc structure but have ignored them in the hcp structure; see Sec.~\ref{sechcpstructure}. Therefore, I cannot exclude that the calculated maxima of $T_{c}$ in the hcp structure are considerably shifted in reality. At this stage, I only can state that a pure $s$-$d$ band structure in hcp solid solutions leads to positions of the maxima which are similar to what is found in bcc materials. 

\acknowledgements{I wish to thank Ernst Helmut Brandt for stimulating discussions and his great help in completing this work. I thank Max Wagner and G\"unther Tatzel for critical comments on the manuscript and Uwe E\ss mann for continuing support.}

\renewcommand{\theequation}{\Alph{section}\arabic{equation}}

\begin{appendix}

\ab{General}
\label{general}
Consider a metal with the space group $G$, the point group $G_{0}$, and $\mu$ atoms in the unit cell. The positions of the atoms are written as
\ag
\vec T = \vec R + \vec\varrho_{i},
\label{B1}
\eg
where $\vec R$ and $\vec\varrho_{i}$ ($i = 1$ to $\mu$) denote the lattice points and the positions of the $i$th atom within the unit cell, respectively. The elements
\ag
a = \{\alpha |\vec t\}
\label{B2}
\eg
of $G$ consist of a point group operation $\alpha$  and a translation
\ag
\vec t = \vec \tau (\alpha ) + \vec R,
\label{B3}
\eg
which is the sum of the nonprimitive translation $\vec \tau (\alpha )$ associated with $\alpha$ and a primitive translation $\vec R$. 

The operators $P(a)$ act on a function of position, $f(\vec r)$, according to
\ag
P(a)f(\vec r) = f(\alpha^{-1}\vec r - \alpha^{-1}\vec t ),
\label{B5}
\eg
and on Pauli's spin functions,
\ag
u_{s}(t) = \delta_{st},
\label{B6}
\eg
according to
\ag
P(\alpha )u_{s}(t) = \sum_{s'} d_{s's}(\alpha )u_{s'}(t).
\label{B7}
\eg
The matrices $[d_{s's}(\alpha )]$ are the representatives of the two-dimensional double-valued representation $D_{1/2}$ of the three-dimensional rotation group $O(3)$, and $s = \pm \frac{1}{2}$ and $t = \pm \frac{1}{2}$ denote the spin quantum number and the spin coordinate, respectively.

The effect of $K$ is given by the equations\cite{bc}
\ag
Kf(\vec r) = f^*(\vec r)
\label{effkf}
\eg
and
\ag
Ku_s(t) = g_su_{-s}(t),
\label{symkpauli}
\eg
with\cite{streitwolf}
\ag
g_{\pm 1/2} = \mp i.
\label{gspauli}
\eg

\ab{Superconducting band complexes}
\label{defsigmaband}
The Bloch functions belonging to a $\sigma$-band complex can be unitarily transformed into spin dependent Wfs which are best localized, symmetry adapted to the paramagnetic group 
\ag
M^P = G + KG,
\label{pg}
\eg
and situated at the atoms in the sence that there is at each atom the center of symmetry of exactly one Wannier function. $G$ denotes the space group and $K$ stands for the operator of time inversion. A $\sigma$-band complex contains just as many bands as there are atoms in the unit cell.

In this section, I shall give the group-theoretical condition to identify such a $\sigma$-band complex in a given calculated band structure. 

Consider a set of $\mu$ bands in the given metal and assume the Bloch functions in the points of symmetry $P_{\vec k}$ to be labeled by double-valued $2\mu$-dimensional representations $R^{d}_{\vec k}$ of the group $G_{\vec k}$ of $P_{\vec k}$. $R^{d}_{\vec k}$ may be irreducible or is the direct sum of irreducible double-valued representations. This set of $\mu$ bands forms a $\sigma$-band complex if two conditions are fulfiled.\\
(1) In a $\sigma$-band complex, only representations $R^{d}_{\vec k}$ occur which can be written as a Kronecker product of a single-valued representation $R_{\vec k}$ with the representation $D_{1/2}$,
\ag
R^{d}_{\vec k} = D_{1/2}\times R_{\vec k}.
\label{B8}
\eg
Again, $R_{\vec k}$ may be irreducible or is the direct sum of irreducible single-valued representations.\\
(2) In a $\sigma$-band complex, the matrix representatives $D_{\vec k}(a)$ of the representation $R_{\vec k}$ in Eq. \gl{B8} satisfy, at each symmetry point $P_{\vec k}$, the equation
\aga
\lefteqn{\mbox{trace }D_{\vec k}(a)}\nonumber\\ 
&=& d(\alpha )e^{-i\alpha\vec k\cdot\vec t}
\sum_{i = 1}^{\mu}n_{i}(a)e^{-i\vec \varrho_{i}\cdot (\vec k - \alpha\vec k)} \mbox{ for }a \in G_{\vec k},
\label{B9}
\ega
where $d(\alpha )$ stands for the representatives of any real one-dimensional single-valued representation, say $\Gamma_{0}$, of $G_{0}$ and
\ag
n_{i}(a) = \left\{
\begin{array}{r l}
1 & \mbox{if } \alpha\vec \varrho_{i} + \vec t = \vec \varrho_{i} + \vec R\\
0 & \mbox{else}
\end{array}
\right.
\label{B10}
\eg
with $\vec R$ being a primitive translation (which, of course, may be equal zero).\\ 
Eq.~\gl{B9} is derived from Eqs. (1.8) and (4.28) of Ref.~\onlinecite{ew1} (cf. Sec. 3 of Ref.~\onlinecite{ew2}) assuming that there is one Wannier function with $\Gamma_{0}$ symmetry at each atom. Eq.~\gl{B9} is satisfied for all $a \in G_{\vec k}$ if it is satisfied for one representant of each class of $G_{\vec k}$.

\ab{spin dependent Wannier functions}
\label{spindependentwfs}
In this section the spin dependent Wfs belonging to $\sigma$-band complexes will be defined.
Consider a $\sigma$-band complex consisting of $\mu$ bands. The Bloch functions $\varphi_{\vec kq}(\vec r)$ of this band complex can be unitarily transformed into spin dependent Wfs
\aga
\lefteqn{w_{im}(\vec r - \vec R - \vec\varrho_{i}, t)}
\nonumber\\
&=&\frac{1}{\sqrt{N}}\sum_{\vec k}^{BZ}\sum_{q = 1}^{\mu}
e^{-i\vec k\cdot (\vec R + \vec\varrho_{i})}g_{iq}(\vec k)\phi_{\vec kqm}
(\vec r,t),
\label{C5}
\ega
where the spin dependent Bloch functions $\phi_{\vec kqm}(\vec r,t)$ have $\vec k$-dependent spin directions,
\ag
\phi_{\vec kqm}(\vec r,t) = \sum_{s = -\frac{1}{2}}^{+\frac{1}{2}}
f_{sm}(q,\vec k)u_{s}(t)\varphi_{\vec kq}(\vec r).
\label{C6}
\eg
The first sum in Eq.~\gl{C5} runs over the $N$ wave vectors of the Brillouin zone (BZ), $q = 1$ to $\mu$ denotes the band index, the $\mu$-dimensional matrices $[g_{iq}(\vec k)]$ are, for each $\vec k$, unitary, and the two-dimensional matrices $[f_{sm}(q,\vec k)]$ are, for each q and $\vec k$, unitary. $i = 1$ to $\mu$ labels the Wfs at different atoms in the unit cell and the crystal-spin quantum number $m = \pm\frac{1}{2}$ distinguishes between the functions at the same atom.

In a $\sigma$-band complex, that means, if Eqs.~\gl{B8} and \gl{B9} are satisfied, the matrices $[g_{iq}(\vec k)]$ and $[f_{sm}(q,\vec k)]$ can be chosen in such a way\cite{ew1,ew2,ew3} that\\
(1) the spin dependent Bloch functions $\phi_{\vec kqm}(\vec r,t)$ vary smoothly through the whole $\vec k$ space and\\
(2) the Wfs are symmetry adapted to $M^P$ according to
\aga
\lefteqn{P(a)w_{\vec Tm}(\vec r,t)}\nonumber\\
&=& d_{i}(\alpha )\sum_{m' = -\frac{1}{2}}^{+\frac{1}{2}}
d_{m'm}(\alpha )w_{\vec T'm'}(\vec r, t) \mbox{ for } a \in G
\label{C7}
\ega
and
\ag
Kw_{\vec Tm}(\vec r,t) = \pm w_{\vec T -m}(\vec r,t),
\label{C9}
\eg
with the abbreviations
\ag
w_{\vec Tm}(\vec r,t) \equiv w_{im}(\vec r - \vec R - \vec\varrho_{i}, t)
\label{C8}
\eg
and
\ag
\vec T' = \alpha (\vec R + \vec\varrho_{i}) + \vec t.
\label{C10}
\eg
The plus in Eq.~\gl{C9} is defined to belong to $m = +\frac{1}{2}$ and the minus to $m = -\frac{1}{2}$. As a consequence of Eq.~\gl{B8}, the two-dimensional matrices $[d_{m'm}(\alpha )]$ belong to the representation $D_{1/2}$ of $O(3)$ and the coefficients $d_{i}(\alpha )$ are c-numbers with $|d_{i}(\alpha )| = 1$. The smoothness of the spin dependent Bloch functions $\phi_{\vec kqm}(\vec r,t)$ guarantees that the spin dependent Wfs are optimally localizable.

[If in Eq.~\gl{B10} $n_{i}(\alpha ) = 1$, the c-numbers $d_{i}(\alpha )$ are equal to $d(\alpha )$ as given in Eq. \gl{B9}. In all cases we have $d_{i}(\alpha ) = \bar{D}_{ji}(\alpha)$ with $j$ labeling the non-vanishing element in the  $i$th column of the matrix $[\bar{D}_{ji}(\alpha)]$ given in Eq. (2.18) of Ref.~\onlinecite{ew2} which has only one non-vanishing element in each column.]

If in Eq.~\gl{C6} we have
\ag
f_{sm}(q,\vec k) = \delta_{sm},
\label{C11}
\eg
the Wfs are usual Wfs. We have spin dependent Wfs if the matrix $[f_{sm}(q,\vec k)]$ is not independent of $\vec k$. 

Usual Wfs could be constructed only if the $\sigma$-band complex would not be connected with other bands (if we still demand that the Wfs are best localized and symmetry adapted).\cite{ew1,ew2} In the metals, however, any energy band is connected which other bands at the point and lines of symmetry. Therefore, the matrix $[f_{sm}(q,\vec k)]$ cannot be chosen independent of $\vec k$. 

The abbreviation in Eq.~\gl{C8} is possible since there is one Wannier function at each atom in a $\sigma$-band complex. Therefore, the description of the nonadiabatic Heisenberg model given in Sec. II of part I of this paper need not be changed in the case of $\sigma$-band complexes consisting of more than one band.

\end{appendix}

\end{multicols}
\widetext
\begin{center}
\begin{tabular}{p{40pc}}
\hline
\hline\\
\end{tabular}
\end{center}
\begin{multicols}{2}
\narrowtext

\begin{table}[p] 
\begin{center}

\begin{tabular}{cc}

\begin{tabular}{cc}
\multicolumn{2}{c}{(a)}\\
\hline
1 & ( 0, 0, 0)\\
2 & ( 0, 0, 1)\\
3 & ( 1, 0, 0)\\
4 & ( 0,-1, 1)\\
5 & ( 0, 1,-1)\\
6 & ( 0,-1, 0)\\
\hline
\end{tabular}
&\@{\qquad}
\begin{tabular}{ccp{.2cm}cc}
\multicolumn{5}{c}{(b)}\\
\hline
1 & ( 0, 0, 0)&&7&( 1, 0, 0)\\
2 & ( 0, 0, 1)&&8&( 1, 0, 1)\\
3 & ( 0, 1, 0)&&9&(-1, 0, 0)\\
4 & ( 0, 1, 1)&&10&(-1, 0, 1)\\
5 & (-1, 1, 0)&&11&( 1,-1, 0)\\
6 & (-1, 1, 1)&&12&( 1,-1, 1)\\
\hline
\end{tabular}\\
\end{tabular}
\end{center}
\caption{The points $M$ of the reciprocal lattice related, according to Eq.~\gl{2.1}, to the $i$th free-electron energy band given in Fig.~\ref{figure3} and~\ref{figure4}, respectively. Each column lists $i$ together with $\vec K_{M}$ in the basis given in Table 3.3 of Ref.~\protect\onlinecite{bc}.\\ 
(a) bcc structure. (b) hcp structure.}
\label{table2}
\end{table} 

\begin{table} 
\begin{center}
\begin{tabular}{c}
\hline
$\Gamma_{1} + \Gamma_{12} + \Gamma_{25}'$\\
$H_{1} + H_{12} + H_{25}'$\\
$P_{1} + P_{3} + P_{4}$\\
$3N_{1} + N_{2} + N_{3} + N_{4}$\\
\hline
\end{tabular}
\end{center}
\caption{Representations of an ideal set of $s$-$d$ bands in the bcc structure with notations given in Sec. A3-3 of Ref.~\protect\onlinecite{sl}.}
\label{table1}
\end{table}

\begin{table}
\begin{center}
\begin{tabular}{c}
\hline
$2\Gamma_{1}^{+} + 2\Gamma_{4}^{-} + \Gamma_{5}^{+} + 
\Gamma_{5}^{-} + \Gamma_{6}^{+} + \Gamma_{6}^{-}$\\
$2H_{1} + H_{2} + 3H_{3}$\\
$K_{1} + K_{2} + K_{3} + K_{4} + 3K_{5} + K_{6}$\\
$2A_{1} + 2A_{3}$\\
$4L_{1} + 2L_{2}$\\
$3M_{1}^{+} + M_{1}^{-} + M_{2}^{+} + 3M_{2}^{-} +
M_{3}^{+} + M_{3}^{-} + M_{4}^{+} + M_{4}^{-}$\\
\hline
\end{tabular}
\end{center}
\caption{Representations of an ideal set of $s$-$d$ bands in the hcp structure [as determined by means of Eq.~\protect\gl{B9}, see Sec.~\protect\ref{sechcpstructure}] with notations given in Sec. A3-1 of Ref.~\protect\onlinecite{sl}.}
\label{table3}
\end{table}

\begin{table}
\begin{center}
\begin{tabular}{cp{5pt}cp{5pt}cp{5pt}c}
\hline
$\Gamma_{6}^{+}$ && $H_{6}^{+}$ && $P_{6}$ && $N_{5}^{+}$\\
$\Gamma_{7}^{+}$ && $H_{7}^{+}$ && $P_{7}$ && $N_{5}^{+}$\\
$\Gamma_{6}^{-}$ && $H_{6}^{-}$ && $P_{6}$ && $N_{5}^{-}$\\
$\Gamma_{7}^{-}$ && $H_{7}^{-}$ && $P_{7}$ && $N_{5}^{-}$\\
\hline
\end{tabular}
\end{center}
\caption{Double-valued representations $R^{d}_{\vec k}$ of the four superconducting bands in the bcc structure with notations of Ref.~\protect\onlinecite{elliott}.}
\label{table4}
\end{table}

\end{multicols}
\widetext

\begin{table}
\begin{center}
\begin{tabular}{cp{5pt}cp{5pt}cp{5pt}c}
\hline
$2\Gamma_{6}^{+}$ && $R_{6}^{+} + R_{7}^{-}$ && $M_{6}^{+} + M_{7}^{+}$ && 
$X_{6}^{+} + X_{6}^{-}$\\
$2\Gamma_{7}^{+}$ && $R_{7}^{+} + R_{6}^{-}$ && $M_{6}^{+} + M_{7}^{+}$ && 
$X_{7}^{+} + X_{7}^{-}$\\
$2\Gamma_{6}^{-}$ && $R_{7}^{+} + R_{6}^{-}$ && $M_{6}^{-} + M_{7}^{-}$ && 
$X_{6}^{+} + X_{6}^{-}$\\
$2\Gamma_{7}^{-}$ && $R_{6}^{+} + R_{7}^{-}$ && $M_{6}^{-} + M_{7}^{-}$ && 
$X_{7}^{+} + X_{7}^{-}$\\
\hline
\end{tabular}
\end{center}
\caption{Double-valued representations $R^{d}_{\vec k}$ of the four superconducting band complexes (of two bands each) in the CsCl structure with notations of Ref.~\protect\onlinecite{elliott}. The CsCl structure has the cubic space group $O^{1}_{h}$.}
\label{table5}
\end{table}

\begin{table}
\begin{center}
\begin{tabular}{cp{5pt}cp{5pt}cp{5pt}cp{5pt}c}
\hline
$2\Gamma_{6}^{+} + \Gamma_{6}^{-}$ && $2Z_{6}^{+} + Z_{6}^{-}$ && 
$2X_{5}^{+} + X_{5}^{-}$ && $2P_{6} + P_{7}$ &&
$2N_{3}^{+} + 2N_{4}^{+} + N_{3}^{-} + N_{4}^{-}$\\
$2\Gamma_{7}^{+} + \Gamma_{7}^{-}$ && $2Z_{7}^{+} + Z_{7}^{-}$ && 
$2X_{5}^{+} + X_{5}^{-}$ && $P_{6} + 2P_{7}$ &&
$2N_{3}^{+} + 2N_{4}^{+} + N_{3}^{-} + N_{4}^{-}$\\
$\Gamma_{6}^{+} + 2\Gamma_{6}^{-}$ && $Z_{6}^{+} + 2Z_{6}^{-}$ && 
$X_{5}^{+} + 2X_{5}^{-}$ && $P_{6} + 2P_{7}$ &&
$N_{3}^{+} + N_{4}^{+} + 2N_{3}^{-} + 2N_{4}^{-}$\\
$\Gamma_{7}^{+} + 2\Gamma_{7}^{-}$ && $Z_{7}^{+} + 2Z_{7}^{-}$ && 
$X_{5}^{+} + 2X_{5}^{-}$ && $2P_{6} + P_{7}$ &&
$N_{3}^{+} + N_{4}^{+} + 2N_{3}^{-} + 2N_{4}^{-}$\\
\hline
\end{tabular}
\end{center}
\caption{Double-valued representations $R^{d}_{\vec k}$ of the four superconducting band complexes (of three bands each) in the MoSi$_{2}$ structure with notations given in Table 6.14 of Ref.~\protect\onlinecite{bc}. The MoSi$_{2}$ structure has the tetragonal space group $D^{17}_{4h}$.}
\label{table6}
\end{table}

\begin{table}
\begin{center}
\begin{tabular}{cp{5pt}cp{5pt}cp{5pt}cp{5pt}cp{5pt}c}
\hline
$\Gamma_{7}^{+} + \Gamma_{8}^{-}$ && $H_{5} + H_{7} + H_{9}$ && 
$K_{8} + K_{9}$ && $A_{6}$ &&
$L_{3} + L_{4}$ && $M_{5}^{+} + M_{5}^{-}$\\
$\Gamma_{7}^{-} + \Gamma_{8}^{+}$ && $H_{4} + H_{6} + H_{8}$ && 
$K_{7} + K_{9}$ && $A_{6}$ &&
$L_{3} + L_{4}$ && $M_{5}^{+} + M_{5}^{-}$\\
\hline
\end{tabular}
\end{center}
\caption{Double-valued representations $R^{d}_{\vec k}$ of the two superconducting band complexes (of two bands each) in the hcp structure with notations given in Ref.~\protect\onlinecite{elliott}. }
\label{table7}
\end{table}


\begin{thebibliography}{aaa}
\bibitem{misprint}
In Table XI of Ref.~\onlinecite{elliott} there is a misprint. We have $D_{1/2}\times\Gamma^{\pm}_{5} = \Gamma^{\pm}_{8} + \Gamma^{\pm}_{9}$ and $D_{1/2}\times\Gamma^{\pm}_{6} = \Gamma^{\pm}_{7} + \Gamma^{\pm}_{9}$ in the notation of Sec. A3-1 of Ref.~\onlinecite{sl}.
\end{thebibliography}
\end{document}